\documentclass[twoside]{article} 
\def\gsim{~\rlap{$>$}{\lower 1.0ex\hbox{$\sim$}}}
\def\lsim{~\rlap{$<$}{\lower 1.0ex\hbox{$\sim$}}}
\def\etal{{\it et 
al.\thinspace}}

\usepackage{url}                

\usepackage{graphicx}			

\usepackage[numbers]{natbib}	
\usepackage{lpscabs}			

\begin{document}


\titlearea{Tidal Effects on the Habitability of Exoplanets: The Case of GJ\,581\,d\\}{Rory
  Barnes$^{1,2}$, Brian Jackson$^{3,4}$, Ren\'e Heller$^5$, Richard
  Greenberg$^6$, Sean N. Raymond$^7$, $^1$Department of Astronomy,
  University of Washington, Seattle, WA, 98195-1580,
  rory@astro.washington.edu, $^2$Virtual Planetary Laboratory,
  $^3$Planetary Systems Laboratory, Goddard Space Flight Center, Code
  693, Greenbelt, MD 20771, $^4$NASA Postdoctoral Program Fellow,
  $^5$Hamburger Sternwarte, University of Hamburg, Gojenbergsweg 112,
  21029 Hamburg, Germany, $^6$Lunar and Planetary Laboratory,
  University of Arizona, Tucson, AZ 85721, $^7$Laboratoire d'Astrophysique de Bordeaux (CNRS; Universit\'e Bordeaux I) BP 89, F-33270 Floriac, France}

%


%



Tides may be crucial to the habitability of exoplanets. If such
planets form around low-mass stars, then those in the circumstellar
habitable zone will be close enough to their host stars to experience
strong tidal forces. Tides may result in orbital decay and
circularization, evolution toward zero obliquity, a fixed rotation
rate (not necessarily synchronous), and substantial internal heating
[1--4]. Due to
tidal effects, the range of habitable orbital locations may be quite
different from that defined by the traditional concept of a habitable
zone (HZ) based on stellar insolation, atmospheric effects, and liquid
water on a planet's surface.  Tidal heating may make locations within
the traditional HZ too hot, while planets outside the traditional zone
could be rendered quite habitable due to tides.

\begin{figure}
\begin{center}
\includegraphics[width=\columnwidth]{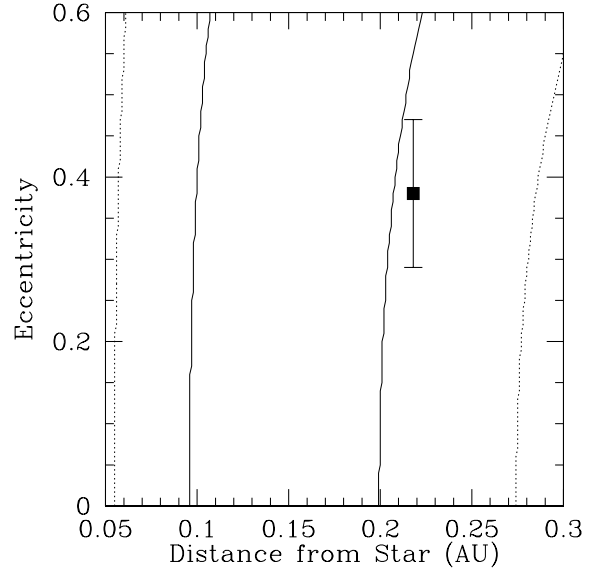}
\caption[]{\label{fig:gj581d}Insolation limits to the
  habitability of GJ\,581\,d. Solid curves correspond to the 0\% cloud
  cover HZ model of [1], dotted curves assume 100\%
  cloud coverage. GJ\,581\,d is the black square, with the 1-$\sigma$
  uncertainty in eccentricity also shown (the errors in $a$ are
  negligible.) }
\end{center}
\end{figure}

Consider for example GJ\,581\,d, a planet with a minimum mass of 7 Earth
masses, a semi-major axis $a$ of 0.22\,AU, and an eccentricity $e$ of
$0.38 \pm 0.09$ ([5]; revised from $a = 0.25$\,AU in [6]). The 
circumstellar habitable zone of [1],
which is a synthesis of [7--8],  predicts this planet receives enough 
insolation to
permit surface water, albeit with some cloud coverage, see 
Fig.~\ref{fig:gj581d}. The small value of $a$ and large value of $e$
suggest that tides may be important, and their potential effects
must be taken into consideration. Given the recent revision of its orbit
[5], we examine the habitability of this planet in the
context of tides. As more planets in the circumstellar HZ of low mass
stars are discovered, a similar analysis should be applied.

\textbf{Rotation Rate} The rotation rate of the planet was tidally
locked in less than 1 Gyr. Tidal locking, however, does not mean the
planet is rotating synchronously, instead it follows the relation
\begin{equation}
\label{eq:spin}
\Omega_\mathrm{eq} = n(1 + ke^2),
\end{equation}
where $\Omega_\mathrm{eq}$ is the equilbrium rotation frequency, $n$ is the
mean motion, and $k$ is a prameter that is
dependent on the tidal model. If tidal bulges lag by a constant phase,
$k = 9.5$ [9,1], but if they lag by a
constant time, then $k = 6$ [10]. Therefore, GJ
581 d may rotate faster than synchronous, with a period of perhaps
about half the orbital period of 66.8 days.

\textbf{Obliquity} Tidal evolution tends to drive obliquities to 0 or
$\pi$ (depending on initial conditions). For GJ\,581\,d, the time for
this ``obliquity locking'' to occur is $\sim 100$\,Myr [4,11]. Should this 
locking occur, the
habitability of GJ\,581\,d may be in jeopardy, even if it is in the
circumstellar HZ, as the poles become a cold trap and can eventually
freeze out the atmosphere [12]. However, perturbations from other planets 
may drive a
chaotic obliquity evolution [13]. For this to occur the orbits of
the other planets in the system must be inclined relative to GJ\,581
d's orbit.

Large mutual inclinations in the GJ\,581 system are likely. An earlier
phase of planet-planet scattering [14] is
evidenced by GJ\,581\,d's large eccentricity, as protoplanetary disk
phenomena are unlikely to produce values larger than 0.3 [15]. Such 
scattering would have likely driven large relative
inclinations ($\gsim 30^\circ$) between planets [14]. So, while tides are 
driving planet d's obliquity
toward 0 or $\pi$, interactions with other planets are preventing this
situation from occuring. Note that the orbital oscillations occur on
$\sim 10^3$ year timescales, which is orders of magnitude shorter than
the obliquity locking timescale. Whether these obliquities
oscillations from the other planets preclude d's habitability is
another matter.

\textbf{Orbital Evolution} The GJ\,581 system is estimated to be 8 Gyr
old [7]. Therefore tides may have played a role in its orbital
history. Tides tend to circularize and shrink orbits with time
[16]. Although these effects are operating on GJ\,581
d, they have resulted in minimal evolution: GJ\,581\,d has always been
in the circumstellar habitable zone. If we assume standard mass-radius
relationships for terrestrial planets [17], then GJ\,581\,d has not drifted an appreciable amount in the last 8 Gyr.

\textbf{Internal Heating} Plate tectonics may be necessary for 
habitability [17]. On Earth, the internal energy to drive this process 
comes from endogenic sources: radioactive decay and energy from formation. 
The sources combine to provide a current heat flux of 0.08 W m$^{-2}$ 
[18]. This value is close to the lower limit for plate tectonics, 
0.04\,W\,m$^{-2}$ derived by [19]. We use their example to 
make a crude estimate of endogenic heat flux on GJ\,581\,d, assuming an age 
of 8 Gyr [6]. They assumed an exponential cooling law:

\begin{equation} 
\label{eq:cool} h_\mathrm{end} = h_\mathrm{end,0}R_\mathrm{p}\rho_\mathrm{p}e^{-\lambda t},
\end{equation}

\noindent
where $h_\mathrm{end}$ are the radiogenic and primordial heating flux (in W\,m$^{-2}$), $h_\mathrm{end,0}$ is a proportionailty constant, $\rho_\mathrm{p}$ is the 
planetary density, $\lambda$ is the the reciprocal of the half-life, and 
$t$ is the age of the system. As a first estimate, [19] set 
$\lambda = 1.5 \times 10^{-10}$, corresponding to the half-life of 
$^{238}$U. The actual cooling times and initial radiogenic inventory of GJ 
581 d could be very different, and Eq.~(\ref{eq:cool}) should be 
considered an order of magnitude estimate. Scaling from the Earth, the 
heat flux from non-tidal sources on GJ\,581\,d is 0.12\,W\,m$^{-2}$, about 3 
times larger than the tectonics limit. Given the uncertainties in this 
calculation, plate tectonics is not a given on GJ\,581\,d.

Perhaps tidal heat, similar to Io's, can provide additional energy.
Tidal heating $H$ inside a planet is equal to the change in orbital energy:
\begin{equation}
\label{eq:heat}
H = \frac{63}{4}\frac{(GM_*)^{3/2}M_*R_\mathrm{p}^5}{Q'_p}a^{-15/2}e^2
\end{equation}
[20--21]. However, in order to assess the
surface effects of tidal heating on a potential biosphere, we can
consider the heating flux, $h = H/4\pi R_\mathrm{p}^2$, through the planetary
surface.  The tidal heat flux of GJ\,581\,d, assuming best fit parameters and
planetary tidal dissipation parameter $Q' = 500$, is 0.01 W m$^{-2}$
[3], about 4 times too low for plate tectonics, and perhaps an order of 
magnitude lower than the endogenic heat flux. Therefore tidal heating, which is also uncertain, could provide a signficant heat source for this planet. Perhaps a combination of endogenic and tidal heat drive plate tectonics, facilitating habitability.

We conclude that tidal effects are an important part of assessing GJ
581 d's potential habitability. 
As more plausibly terrestrial planets are discovered,
these tidal issues need to be applied to them as well in order to
assess their potential habitability.

\vspace{0.2cm}
\bibliography{bibliography}
\noindent[1] Barnes, R. \etal 2008, AsBio, 8, 557. [2] Jackson, B. \etal 
2008, 
MNRAS, 391, 237. [3] Barnes, R. \etal 2009, ApJ, 700, L30. [4] Barnes, R. \etal 2010, in Conference Proceedings to \textit{Pathways Towards Habitable Planets}, Eds: 
D. Gallindo \& I. Ribas, in press. [5] Mayor, M. \etal 2009. A\&A, 507, 
487. [6] Udry, S. \etal 2007, A\&A 469, L43. [7] Selsis, F. \etal 2007, 
A\&A, 476, 137. [8] Williams, D. M. \& Pollard, D. 2002, Int J. AsBio, 1, 61. [9] Goldreich, P. 
1966, AJ, 71, 1. [10] Ferraz-Mello, S. \etal 2008, CeMDA, 101, 171.	
[11] Levrard, B. \etal 2007, A\&A, 462, L5. [12] Selsis, F., personal 
communication. [13] Atobe, K., \& Ida, S. 2007, Icarus, 188, 1. Marzari, 
F. \& Weidenschilling, S. 2002, Icarus, 156, 570. [15] D'Angelo, G. 
\etal 2003 ApJ, 586, 540. [16] Jackson, B. \etal 
2008, ApJ, 678, 1396. [17] Sotin, C. \etal 2007, Icarus, 191, 337. [18] 
Davies, G. 1999, Dynamic Earth (Cambridge, UK; Cambridge UP) [19] Walker, J. C. G., \etal 1981, JGR, 86, 9776. [20] Williams, D. M. \etal 1997, Nature, 
385, 234. [20] Peale, S. J. \etal 1979, Science, 203, 892. 19. [21] 
Jackson, B. \etal 2008, ApJ, 681, 1631.

\bibliographystyle{unsrtnat}

\end{document}